\documentclass[magazine,comsoc]{IEEEtran}
\usepackage[T1]{fontenc}% optional T1 font encoding
\usepackage{comment}
\usepackage{blindtext}

\ifCLASSINFOpdf
\else
\fi
\usepackage{amsmath}
\usepackage[cmintegrals]{newtxmath}
\usepackage{hyperref}

\usepackage{comment}
\usepackage{enumitem}
\usepackage{epsfig}
\usepackage{xcolor}

\begin{document}
% Line-breaks \\ can be used within to get better formatting as desired.

%\title{Bare Demo of IEEEtran.cls for\\ IEEE Communications Society Journals}

\title{D$^3$S: A Framework for Enabling Unmanned Aerial Vehicles as a Service}


\begin{comment}
\author{Farid~Nait-Abdesselam, %~\IEEEmembership{SM,~IEEE,}
        Ahmad~Alsharoa, %~\IEEEmembership{Member,~IEEE,}
        Mohamed~Selim, %~\IEEEmembership{Member,~IEEE,}
        Daji~Qiao, %~\IEEEmembership{SM,~IEEE,}
        and~Ahmed~E.~Kamal%,~\IEEEmembership{Fellow,~IEEE}% <-this % stops a space
%\thanks{M. Shell was with the Department of Electrical and Computer Engineering, Georgia Institute of Technology, Atlanta, GA, 30332 USA e-mail: (see http://www.michaelshell.org/contact.html).}% <-this % stops a space
}
\end{comment}

\author{
        \IEEEauthorblockN
            {
	        Farid Na\"it-Abdesselam\IEEEauthorrefmark{1}
	        Ahmad~Alsharoa\IEEEauthorrefmark{2}
	        Mohamed~Selim\IEEEauthorrefmark{3}
	        Daji~Qiao\IEEEauthorrefmark{3}
	        Ahmed~E.~Kamal\IEEEauthorrefmark{3}\\
	        }
	        \IEEEauthorblockA{\IEEEauthorrefmark{1}University of Missouri Kansas City, USA\\}
	        \IEEEauthorblockA{\IEEEauthorrefmark{2}Missouri University of Science and Technology, USA\\}
	        \IEEEauthorblockA{\IEEEauthorrefmark{3}Iowa State University, USA}
        }

\maketitle

\begin{abstract}
\begin{comment}
\marginpar{\small{I felt the abstract is too long, and I cut some parts of it. It still needs to be reduced further.}}
\end{comment}
%Unmanned Aerial Vehicles (UAVs) take different forms like helicopters, quadcopters, fixed wing planes, blimps, and balloons. UAVs have been used in military applications for quite  some time, and more recently they have emerged also in civilian applications.
%For example, aside from amateur use of drones, Google sent their balloons to help restore communications coverage in Puerto Rico after it was hit by Hurricane Maria. Amazon also is working on the Prime Air project to deliver orders from Amazon.com using drones, and many other applications.
%Unmanned Aerial Vehicles (UAVs) can be used to support wireless connectivity, such as providing wireless service after failures of wireless network components or after disasters, or to simply provide additional bandwidth on demand.
%In this paper, we consider the use of UAVs to provide this type of wireless connectivity service, hence the concept of UAVs as a service (UaaS) for end-to-end connectivity.
In this paper, we consider the use of UAVs to provide wireless connectivity services, for example after failures of wireless network components or to simply provide additional bandwidth on demand, and introduce the concept of UAVs as a service (UaaS).
%\textcolor{red}{Providing continuous coverage, after failures of wireless network components, or additional bandwidth, in situations where there is high traffic demands, is crucial. By introducing the concept of \emph{UAVs as a Service (UaaS)}, in this paper we consider the use of Unmanned Aerial Vehicles (UAVs) to provide wireless communication and networking services, required by such civilian applications, to achieve the end-to-end connectivity for end-users.}
To facilitate UaaS, we introduce a novel framework, dubbed $D\mathrm{^3}S$, which consists of four phases: \textit{demand}, \textit{decision}, \textit{deployment} and \textit{service}.
The main objectives of this framework is to develop efficient and realistic solutions to implement these four phases.
The technical problems include determining the type and number of UAVs to be deployed,
%dimensioning the number of UAVs, 
and also their final locations (e.g., hovering or on-ground),
%if they are to hover, 
which is important for serving certain applications. These questions will be part of the \textit{decision} phase.
They also include trajectory planning of UAVs when they have to travel between charging stations and deployment locations, and may have to do this several times. These questions will be part of the \textit{deployment} phase.
The \textit{service} phase includes the implementation of the backbone communication and data routing between UAVs, and between UAVs and ground control stations.
%This includes the selection of communications technologies, like RF and FSO.
%Another problem in the \textit{service} phase is the data routing among UAVs to implement a mesh network, which also allows access to the core network, and this will have to be done in three dimensions.

%These functionalities will be developed for two stages of operation: \textit{short-term}, in which agile and fast deploying, albeit short life-time UAVs, will be deployed, and \textit{long-term}, in which slower to deploy, but longer life-time UAVs, are deployed.
%Either one of the two stages can be used alone, or the short-term operation can be followed by the long-term operation, depending on the application and its domain.
\end{abstract}


% \fi
\IEEEpeerreviewmaketitle

\section{Introduction}
%\input{1-intro.tex}

%\section{Unmanned Aerial Vehicle Types}
%\input{2-UAV-types.tex}

%\section{The D$^3$S Framework for UaaS}
\label{sec:framework}
In this section we present our proposed framework for implementing UAVs as a Service (UaaS).
This framework consists mainly of four phases, \textit{Demand, Decision, Deployment} and \textit{Service}, abbreviated as D$^3$S.\\[.1in]
\textbf{Phase 1: D (Demand)}
In the \textit{Demand} phase, the entity requesting a service places a request with a set of high-level parameters that characterize the requested service.
These will include: (1) type of request (disaster recovery, self-healing, etc.), (2) the location(s) (in terms of coordinates) of the event and its coverage area, (3) the bandwidth capacity required at each of the locations, (4) mobility characteristics of users, if any, and (5) the time frame of the requested coverage service.
%A more detailed and accurate specification of these parameters can be defined as described in Table~\ref{tab:parameters}. 
The specification of this request will serve as an entry to the second phase, related to the \textit{Decision}.
\begin{comment}
\begin{table}[h!]
\footnotesize \caption{high-level parameters}
\label{tab:parameters}
\begin{center}
\begin{tabular}{||l|p{6cm}||}
\hline
\bf Parameter & \bf Description \\
\hline \hline
 P1 & xxxxxxxxxxxxxxxxxxxxxxxxxxx. \\
 P2 & xxxxxxxxxxxxxxxxxxxxxxxxxxx. \\
\hline
\end{tabular}
\end{center}
\end{table}
\end{comment}
{~}\\[.1in]
\textbf{Phase 2: D (Decision)} %defining the number of UAVs, locations where they have to be deployed, (RT1)
Based on requests made in the \textit{Demand} phase, the \textit{Decision} phase will determine the types of UAVs to deploy, the optimal number of such UAVs, their precise deployment locations and the bandwidth to be used by their communications equipment.
The UAVs will therefore form a mesh network that will provide the requested service to a set of stationary, and/or eventually mobile, ground devices.
%that must be served at the same time, e.g., for providing coverage due to failures of base stations, or due to high traffic demand. 
As different types of UAVs and different deployment locations (e.g., hovering vs. on-ground) may offer different trade-offs between energy consumption, flying time before the need to be recharged, size of the coverage area, etc, all these factors will be taken into consideration when making decisions.
In addition, other mobile devices or devices that do not need a continuous service, such as sensor devices in a farming field, will also be taken into consideration and linked to the determination of the trajectories taken by UAVs in the \textit{Deployment} phase.\\[.1in]
\textbf{Phase 3: D (Deployment)} Once the types and numbers of UAVs and their future locations are determined in the \textit{Decision} phase, the \textit{Deployment} phase will deal with defining the best trajectories of the different UAVs that have to be deployed. The UAVs will then be dispatched either from the same source location, and therefore will be flying as a swarm towards their deployment locations, or from different source locations, and therefore will fly individually and will be gathered one-by-one to converge towards their deployment locations. Multiple configurations to route these UAVs will be taken into consideration in this phase such that the energy resource will be used optimally.\\[.1in]
\textbf{Phase 4: S (Service)} In this phase, the proper coverage service to achieve end-to-end connectivity will be provided. This includes communication of UAVs with ground users (stationary or mobile), routing of data between UAVs, and routing of data to and from access points to the core network.\\[.1in]
\textbf{Short-Term vs. Long-Term.}
Two time scales will be used to provide service: \textbf{short-term} and \textbf{long-term}.
The short-term service provisioning refers to the use of UAVs that can be deployed with agility, e.g., drones.
These UAVs typically have short flying and hovering times but can be deployed to provide service with a very short delay.
The long-term service provisioning uses UAVs that take longer to deploy, but can stay in service for a long time without requiring maintenance or recharging, e.g., helikites, airships and balloons.
The use of short-term followed by long-term, short-term only, or long-term only, depends on the application, and the application domain and its properties.
For example, disaster recovery and self-healing of wireless systems can use short-term followed by long-term service. For applications involving forecasted increase in bandwidth demand, such as in football games, pre-planning can be implemented ahead of the event and long-term service can be provisioned.
%before the event.
%
The introduction of these two time scales, and the transitioning between them will be implemented by the Decision, the Deployment and the Service phases of the framework.
%RT3+RT4 as one RT3(we try to include the SDWN for the routing planning and reconfiguration with machine learning classifying and predicting new changes in the routing process/service)

%Each of our target applications requires some type of packet routing, either from a UAV to a base station,from a base station to one or more UAVs, or from a base station to another base station through a network of UAVs. In this thrust of the project, we will explore novel methods and protocols for packet routing in the dynamic and challenging environment of a 3D UAV network

\section{Demand Forecasting and Characterization}
\label{sec:dem-forecast}

In the first phase of the framework, the entity requesting service must provide information about the requested service in terms of the type of service, e.g., due to disaster, the devices to be served, whether they are stationary or mobile, and the requested service rates.
The requested duration of service can also be identified, and whether the service is continuous or intermittent, e.g., for sensors.
The information may also be updated with time.

This information can be provided formally as follows:
\begin{itemize}[leftmargin=*]
\item
A set, $\mathcal{D}$, of stationary devices which may include sensors, IoT devices and other stationary communicating devices.
Each of the devices is defined in terms an ordered pair which identifies the location in the two-dimensional Cartesian plane
and the rate requirements of the device.
The information does not have to be for individual devices, but rather for groups of devices.
Each group can be treated collectively as one point of service.
If the requested service rates change, then this information may be updated with time.
%\vspace{-.05in}
\item
A set, $\mathcal{M}(t)$, of mobile devices, e.g., mobile users equipment (UEs), service vehicles, emergency vehicles, etc.
Each mobile UE is defined in terms of an ordered pair which identifies the location in the two-dimensional Cartesian plane
and the rate requirements of the device at time $t$.
Due to mobility, the locations in $\mathcal{M}(t)$ have to be updated with time.

%\vspace{-.05in}
\item
The total bandwidth available for communications which consists of a set, $\mathcal{W}$ of fixed bandwidth channels.
A device in $\mathcal{D}$ or $\mathcal{M}$  may use one or multiple of these channels, depending on the rate requirements, the channel gains between the device and the associated UAV, as well as interference from other UAVs.
\end{itemize}
The rates identified for device service can be regarded as minimum required rates.

The type of service and the requested service duration are also important in planning service and the devices to be committed to the service.
For example, a service due to a disaster is very different from a service due to an increased traffic demand.
In the first case there is no service, and guaranteeing a minimal level of service is important.
In the second case, there is available service, but the network is congested and the requested service is to improve on available one.

\section{Decision and Dimensioning Phase}
\label{sec:dimensioning}

\vspace{-.05in}
In the decision and dimensioning phase, the information collected in the demand phase is used to determine the number of UAVs, their locations and bandwdith assignments such that service can be provided to the sets of stationary and mobile devices $\mathcal{D}$ and $\mathcal{M}$, respectively.
%For the case of mobile devices, trajectory planning will also be important, and this will be treated in \textcolor{red}{Subsection \ref{sec:trajectory}}.
For the sake of illustration, we focus on downlink communications only.
Backhauling is implemented in a distributed manner between UAVs, i.e., using multihop communications to the nearest stationary base station.

\paragraph{Short Term Dimensioning}
To provide service to the set of stationary devices, $\mathcal{D}$, defined above, a subset of the UAVs, $\mathcal{U}$ will act as base stations.
The objective of the UAV dimensioning problem is twofold: 1) to minimize the number of UAVs, and 2) to maximize the operational lifetime of UAVs.
These two objectives may be contradictory  since one may be able to reduce the number of UAVs but they will have to cover wider geographical areas, hence consuming more energy and depleting the UAVs batteries faster. 

Therefore, the dimensioning phase is solved using a dual objective optimization problem:
%\vspace*{-.1in}
\begin{equation}
\mathrm{Minimize} ~~~ (f_U, -f_T ) \label{eqn1}
\end{equation}
where $f_U$ is the number of used UAVs and $f_T$ is a function of the UAVs' lifetimes.
The objective function minimizes $-f_T$, which is equivalent to maximizing $f_T$.
$f_T$ can be expressed as the minimum lifetime over all UAVs, and minimizing $-f_T$ corresponds to maximizing the minimum lifetime over all UAVs.
The lifetime of a UAV depends on the UAV's battery energy available for communications after subtracting the mechanical energy, viz., energy used for flying and hovering.
The UAV's lifetime is obtained by dividing this energy by the power used for communications.  
The mechanical energy used by the UAV to fly from initial to hovering location, and from hovering to charging station, are dependent on the chosen location for the UAV.
The optimal hovering location of UAV $i$ at time $t$, $l_i (t)$, is in the three dimensional Cartesian plane, which may include an altitude of 0, i.e., ground level deployment.

There are two types of communications in which the UAVs are involved, and these influence the use and sharing of the bandwidth: UAV-to-user and UAV-to-UAV communications.
These are captured in the dimensioning phase by using two association matrices: \\
1) the device-UAV association, which is captured using the matrix, $A_{|\mathcal{D}| \times |\mathcal{U}|}$ where each matrix element is a binary variable that is only equal to 1 if device $i\in \mathcal{D}$ uses UAV $j\in \mathcal{U}$.
Typically each device is constrained to use exactly one UAV.
Determining whether UAV $j$ is used or not can be obtained from this matrix, and can also be be used to obtain the number of needed UAVs.
$P_{i,j}$ is the transmission power from UAV $j$ to user $i$.\\
2) If UAVs communicate among themselves using the same RF spectrum, then a symmetric UAV-to-UAV association matrix, $\tilde{A}_{\mathcal{U} \times \mathcal{U}}$ is defined.
A matrix element is 1 if two UAVs communicate.
The power used for communication between them is $\tilde{P}_{mn}$.\\
The dimensioning phase evaluates both $A$ and $\tilde{A}$.
%UAVs $m$ and $n$ can be in Tiers 1 and 2, as explained in the system model in Subsection \ref{sec:model}.

%\begin{equation}
%\sum\limits_{k=1,k\neq j}  \tilde{a}_{kj} \leq %1~~\forall i
%\end{equation}
%If device $i$ is associated with UAV $j$, then the communication power consumed by UAV $j$ to transmit to device $i$, 
$P_{ij}$, $\tilde{P}_{mn}$ and the downlink rate to device $i$ are determined by the dimensioning phase in order to guarantee that the spectrum is shared between these two types of communications to achieve the minimum required rate, even with the presence of interference. 
The interference depends on the channel gain between pairs of devices, and depends on the distances between them.
The backbone rate is also a function of the rate of communications between the UAVs and their served UEs, and is determined by the backbone routing.
In the case of employing OFDMA, interference is not present.

In case the resources are not sufficient to guarantee the minimum required rates for devices, a third objective can be added, and will be the maximum violation of the bit rate among all devices, and this will be minimized.

Solving the optimization problem expressed by objective function (\ref{eqn1}), and the constraints that are formulated based on the above discussion should result in the optimal dimensioning including the number and hovering locations of UAVs, and their association with users as well as their transmission power levels.
However, since the problem is a multi-objective optimization problem, the solution will not give a single solution, but a Pareto front of the non-dominated solutions.  
Solving this problem is not easy due to a number of reasons: 1) it is a dual objective optimization problem, 2) it includes binary variables, which makes the problem NP-hard, and 3) it is highly non-convex.
Therefore, typically approximations and heuristics are employed for solving this optimization problem within a time frame that is suitable for the problem, while producing close-to-optimal solutions.
Solution approaches include device clustering, binary variable relaxation, successive convex approximation and evolutionary programming approaches.

\typeout{
The above is a formulation for a single group of devices.
However, in the applications that we consider there may be several groups of users, and the resources will need to be allocated to serve all of them.
We therefore plan to extend this formulation to handle the case of multiple groups of devices, given the same set of UAVs.
This will introduce a comprehensive strategy for operating the UAVs to provide the required service, and can be used in many application domains, including serving high traffic demands, recovery from disasters, and also communications in real-time IoT environments that do not have a communication infrastructure, like battlefields.
This will also introduce more complexity to the problem, and group-specific objectives may be added to the objective function, hence resulting in a higher dimensional Pareto front.
Since the solutions will be in terms of the Pareto front, a strategy for selecting the operational strategy will need to be developed.  
If groups compete for a service, then we plan to develop games, and possibly auction strategies for the allocation of UAV services to the device groups.
}

\paragraph{Long Term Dimensioning}
Dimensioning for the long term stage is similar to short term dimenioning, except that the characteristics of the UAVs used for long term service are taken into consideration.
Since energy efficient UAVs can stay afloat for a long time, they will need to adapt to changing traffic demands and they may also use high power levels for communications, hence achieving higher rates and covering wider areas.
Transitioning from short term to long term needs to consider the coexistence of UAVs of different types and different capabilities.
The simplest, but not necessarily the most efficient approach is to deploy all long term UAVs, and then withdraw all short term UAVs.

\section{Deployment and Trajectory Planning Phase}
The information from the demand and decision phases play significant role in the deployment and trajectory planning phase. Information from the decision phase such as the rate requirements for certain users can affect the UAVs trajectory. For example, obtaining good channel gains between the UAVs and targeted users require, in general, the UAVs to move closer to the targeted users to obtain better channel, which in turn expect increasing the achievable rate.
On the other hand, the information from the decision phase such as number of UAVs and bandwidth limitation will directly affect the deployment and trajectory design by limiting the available resources to use.

By exploiting a careful trajectory design of the UAVs, significant performance gains can be achieved compared to traditional static wireless systems. However, several limitation factors need to be considered. The first one is the instantaneous battery levels of the UAVs, where each UAV determines its battery level periodically to make sure it has enough battery for both hovering and communications. The second factor is the nearby available charging stations. We assume that each charging station can accommodate a maximum number of UAVs at a time instance. Thus, each UAV needs to pre-define the available charging stations in order to land for charging when needed. The third factor is recharging period, i.e., how much time the UAV needs to stay in the charging station. This depends on the decision of the central unit which is based on the user's demand. Finally, the last main factor is the safely path planning, for example, the UAVs are required to avoid flying over some restricted regions, such as airports and military regions. Also, they are required to respect the obstacles on the way, such as high buildings and avoid collisions with other UAVs.

Let us assume that we have certain number of charging station locations with maximum UAVs that can be accommodated in each charging station. %The coordinates of the charging stations and central units are the same and are given by $C_c=(x_c,y_c,z_c)$, $\forall c=1,..,CH$. 
%Let us assume that we have $\mathcal{CH}=\{CH_1,CH_2,..,CH_{CH} \}$ charging station locations with $U_{ch}$ maximum UAVs that can be accommodated in one charging station. The coordinates of the charging stations and central units are the same and are given by $C_c=(x_c,y_c,z_c)$, $\forall c=1,..,CH$. 
%Thus, the following two conditions %need to be respected
%\begin{equation}\label{cha1}
%\sum\limits_{l=1}^U \%epsilon_{l}^{n}(i) \leq U_{ch}, \%quad \forall i  
%\end{equation}
%\begin{equation}\label{cha2}
%\sum\limits_{l=1}^U %\epsilon_{l}^{n}(i)  \leq 1, \quad %\forall i,c, \quad \text{where} %\quad i \neq C_c.
%\end{equation}
%where constraint~\eqref{cha1} limits
Thus, two constraints need to be respected. First, the maximum number of UAVs that can be charged during each time slot at each charging station.% is limited to no more than $U_{ch}$. %Constraint~\eqref{cha2} prevent %positioning 
Second, no more than one UAV can be at the same location during each time slot.
%These constraints can be expressed using the above variables and parameters.
%%and $C$ center units distributed in target area. 
%We can now  all possible scenarios: 1) when the UAV is located not in %the cahrging station at time $t-1$; 2) when the UAV stays at the same %serving location but not at charging stations $i \notin \mathcal{CH}$; %3) when the UAV moves to return to the charging station (i.e., $i \in %\mathcal{CH}$) while it was located at the serving location $j \notin %\mathcal{CH}$ during time $t-1$; and 4) when the UAV decides to remain %in the charging station $i= \notin \mathcal{CH}$. 
%In Table~\ref{Tab1_tra}, $\boldsymbol{\epsilon^b}$ corresponds to a %binary matrix of size number of UAVs $\times$ number of discrete %locations. Its entries $\epsilon_{l}^{t}(i)$ indicates the location of %the UAV $l$.  In other words, $\epsilon^t_{l}(i)=1$ if UAV $l$ is %placed at location $i$ during time slot $t$, and  %$\epsilon^t_{l}(i)=0$ otherwise. Also, we refer to $P_f, P_s$, and %$P_{ch}$ as the flying, serving, and charging powers, respectively.
Therefore the possible scenarios can be summarized as follows: 1) when the UAV is located not in the charging station;% at time $t-1$; 
2) when the UAV stays at the same serving location but not at charging stations; 3) when the UAV moves to return to the charging station while it was located at the serving location; and 4) when the UAV decides to remain in the charging station.
%In Table~\ref{Tab1_tra}, $\boldsymbol{\epsilon^b}$ corresponds to a binary matrix of size number of UAVs $\times$ number of discrete locations. Its entries $\epsilon_{l}^{t}(i)$ indicates the location of the UAV $l$.  In other words, $\epsilon^t_{l}(i)=1$ if UAV $l$ is placed at location $i$ during time slot $t$, and  $\epsilon^t_{l}(i)=0$ otherwise. Also, we refer to $P_f, P_s$, and $P_{ch}$ as the flying, serving, and charging powers, respectively.

We categorize the ground users into stationary and mobile users. The only difference between these two types of users is that the speed of stationary users is equal to 0.
We assume that the total time period is discretized into equal sub-slots, where the communication channel is approximately unchanged during the sub-slot.
Furthermore, We assume that each user can be associated to one UAV at most during each short time slot.
On the other hand, based on the moving speed of the ground user and UAV, we assume that the maximum distances the ground user and UAV can travel in each sub-slot are limitted by their speeds during the sub-slot.

\section{Service Phase}

In this service phase, the proper coverage service to achieve end-to-end connectivity will be provided. This includes communication of UAVs with ground users (stationary or mobile), routing of data between UAVs, and routing of data to and from access points to the core network. In order to provide UaaS for end-to-end connectivity, it is critically important to establish a reliable backbone network between UAVs to allow reliable, low-latency data delivery either from a UAV to a base station, or from a base station to one or more UAVs, or from a base station to another base station through a network of UAVs.
%We have identified a few packet routing methods that are particularly suitable for the dynamic and challenging environment of UAV networks.
%Researchers have long studied protocols for mobile ad-hoc networks (MANETs), and more recently, vehicular ad-hoc networks (VANETs). Networks of UAVs have recently earned their own distinction, flying ad-hoc networks (FANETs)~\cite{bekmezci:2013survey,2016-Gupta-Survey}. 
%, due to the extreme about of mobility possible and the addition of a third degree of freedom of movement.

Existing works on routing in UAV networks have typically used or adapted classic MANET (Mobile Ad-hoc Networks) protocols. These protocols are classified as either proactive or reactive,
%\cite{rfc3626,perkins1994highly,batman:2008}
%or reactive~\cite{rfc3561,Johnson1996}, 
depending on whether they maintain routes \textit{a priori} or build routes on demand. 
%Well-known proactive protocols include the Optimized Link State Routing protocol (OLSR)~\cite{rfc3626}, the Destination-Sequenced Distance Vector routing protocol (DSDV)~\cite{perkins1994highly}, and the Better Approach to Mobile Adhoc Networking protocol (BATMAN)~\cite{batman:2008}. Reactive protocols include the Ad-hoc On-demand Distance Vector routing protocol (AODV)~\cite{rfc3561} and the Dynamic Source Routing protocol (DSR)~\cite{Johnson1996}. 
Hybrid protocols, such as the Hybrid Wireless Mesh Protocol (HWMP) of the IEEE 802.11s standard, also exist.
%~\cite{hiertz:2010ab}. 
However, these classic protocols generally perform poorly in UAV networks where nodes are moving fast.
%
%The authors of~\cite{pojda:2011ab} experimentally compare HWMP, BATMAN, and OLSR, with performance ranking in that order. However, experiments were performed only with static and slow-moving nodes. \cite{cheng:2012ab} uses simulations to compare OLSR, AODV, and OSPF-MDR (which is similar to OLSR) in a FANET scenario, with OLSR generally performing the best. In~\cite{zafar:2017ab}, a multicluster architecture is proposed and evaluated in simulations using OLSR, DSDV, and AODV, with AODV generally performing poorly.
%
Some works have adapted MANET protocols for use in UAV networks~\cite{8255752}.
%\cite{zheng:2014ab,rosati:2016ab,li:2012ab,alshabatat:2010ab,shi:2012ab,lin:2012ab,li:2012cd}. 
%such as ML-OLSR~\cite{zheng:2014ab}, is a mobility- and load-aware version of OLSR. Predictive OLSR (P-OLSR)~\cite{rosati:2016ab}, uses GPS information to predict link states based on relative speed and direction of UAVs. Other modifications of OLSR specifically for UAV networks include COLSR~\cite{li:2012ab}, and DOLSR~\cite{alshabatat:2010ab}. A cluster-based, location-aided version of DSR for UAV networks is proposed in~\cite{shi:2012ab}. Other UAV network routing protocols include  GPMOR~\cite{lin:2012ab}, a geographic routing protocol that considers mobility and orientation, and RGR~\cite{li:2012cd}, a protocol that combines reactive and geographic routing. 
Although these excellent efforts were successful in handling some of the scenarios for UAV networks, more innovations are needed to improve the reliability and latency performances of the routing protocols.
%scale the routing to many hops in 3D.

%\paragraph{Our Approaches:}
In the following, we describe three methods that may be more suitable for routing in the dynamic and challenging environment of UAV networks:
%Leveraging on the PIs' expertise and previous experiences in ad hoc routing protocols, we propose to study the feasibility of the following routing methods for UAV networks: 
(1) proactive routing based on cohesive swarming and machine learning; (2) fast-converging reactive routing based on back-pressure; and (3) opportunistic routing based on anycast.
%We will first study each of these methods in depth, based on which to integrate all or some of these techniques to become the final routing solution.
\\[.1in]
\textbf{Proactive Routing based on Cohesive Swarming and Machine Learning}:
Unlike conventional wireless ad hoc networks, designing optimal multi-path routing and congestion control algorithms for UAV networks is particularly challenging due to the highly dynamic energy-aware UAV flight maneuvers, which yields constantly changing network topology and fluctuating channel qualities. Classic proactive routing methods are known to perform poorly in such an environment. One possible way to enhance proactive routing for UAV networks is to combine it with cohesive swarming, which coordinates UAVs to form a swarm or shape that suits best the underlying proactive routing method, as well as the events or users of interest. In addition, machine learning techniques can be used for more accurate traffic prediction and thus to enhance in-routing functions among UAVs.
\\[.1in]
\textbf{Fast-Converging Reactive Routing}:
In addition to accurate predictive proactive routing, designing fast-converging reactive routing methods also plays a critical role in UAV networks. In classic reactive methods, queue-length changes are often used as weights in making dynamic routing decisions. Such methods are known to converge slowly. One possible way to improve the convergence speed is to couple queue-length changes with route update from the previous time slot (called momentum). 
Momentum-based reactive routing methods such as the one proposed in~\cite{Liu16:HeavyBall_INFOCOM} could be a good candidate for routing in UAV networks, due to its low-complexity, and its strong performance guarantees in terms of throughput-optimality, delay reduction, and convergence speed.
\\[.1in]
\textbf{Opportunistic Routing based on Anycast}:
Opportunistic routing refers to the practice of making routing decisions dynamically (instead of following pre-determined routes) based on network events and conditions, such as link availability and quality. The opportunistic approach gives nodes multiple options for forwarding a packet and, thus, is particularly suited to UAV networks where the set of a node's neighbors can be constantly changing. The method proposed in~\cite{wymore:2015ab} could be a good candidate for opportunistic routing in UAV networks, which is a cross-layer approach that merges information from both network and link layers to make dynamic routing decisions based on the available links. Moreover, the opportunistic approach may be integrated with proactive or reactive routing methods to further improve the system performance.
\\[.1in]

\section{Case Study: UaaS for Self-healing}

%\subsection{Case Study I: UaaS for Self-healing}
In this section, we present a case study which illustrates the application of the D$^3$S framework.
This case study addresses the failure of one or more Ground BSs (GBSs) and the application of the D$^3$S framework to provide a backup coverage for the failed GBSs. GBSs failures can be classified to short-term and long-term. Short-term failure is defined as the failure that last for a short period of time. This can last for a few minutes or a few hours. The long-term failure can last for a few days.

In our case study and based on different types of UAVs documented in \cite{Bucaille}, rotary-wing drones are proposed to mitigate the short-term failures as they have the important feature of instant deployment. Moreover, the operational power of DBSs is very high which results in a limited flying/service time which is suitable with the short-term. On the other hand, Helikites are proposed to mitigate the long-term failures as they flies at low altitudes and for long periods of time since they are tethered to a continuous source of power.

\begin{figure}
    \centering
%scale=0.5cm    \includegraphics[width=50mm]{Figures/systemmodelFinal.eps}
    \includegraphics[width=80mm]{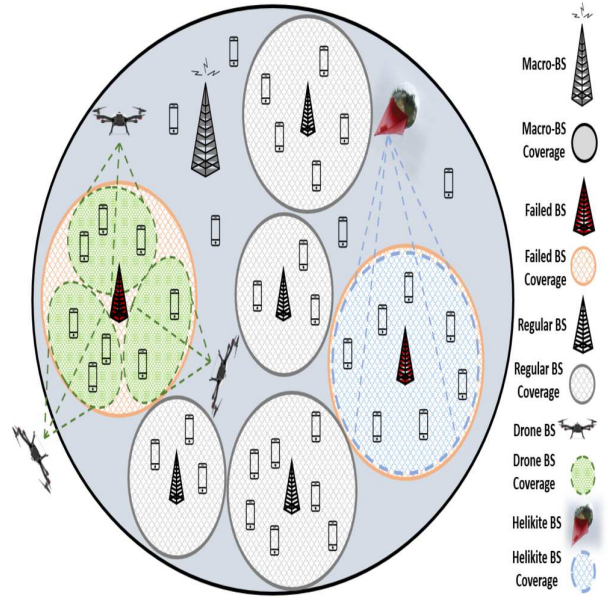}
    \caption{System model during failure.}
    \label{fig:Sysmodel}
\end{figure}

Based on Fig.~\ref{fig:Sysmodel}, the network architecture is based on a heterogeneous network containing a macrocell overlaying a number of GBSs, i.e., small cells. In the presented scenario, two GBSs are failed, one is considered as short-term failure which is healed using Drones and the other is considered as long-term failure where Helikites are used to mitigate the effect of failure.

\subsection{Failure Scenarios}

For the short-term failure scenario, we apply the D$^3$S framework as follows (the results area mainly based on the optimization formulations in reference \cite{8648054} by the authors):
\\[.1in]
\textbf{Phase 1: D (Demand)} Once the network operator detects the failure, a request is placed with a set of parameters related to the failed GBS, i.e., GBS location, coverage area, number of users and the required bandwidth.
\\[.1in]
\textbf{Phase 2: D (Decision)} Based on the detailed previous request and since it is a short-term failure, the decision will be using DBSs to heal this failure. The number of DBSs is decided based on the number of users and requested bandwidth. Based on the given scenario, three DBSs are used to heal the failed GBS. The deployment location is decided based on solving the optimization problem in \cite{8648054}.
\\[.1in]
\textbf{Phase 3: D (Deployment)} In this phase, the deployment depends mainly on the initial locations of the DBSs and the trajectory is determined optimally \cite{8648054}. In the deployed scenario, the initial locations of the DBSs are set such that each GBS is hosting a standby DBS.
\\[.1in]
\textbf{Phase 4: S (Service)} In this phase, we guarantee a minimum achievable rate to the users under the failed GBS. This is always achieved using a rate constraint in the optimization problem \cite{8648054}.\\

For long-term failures, the application of the D$^3$S framwork is exactly the same as the short-term failure except the type of the UAV and the way of deployment. For long-term failures, we use Helikites. Based on the Demand phase, we may use Helikite(s) only (if the application is not time sensitive) or we use DBSs first until deploying the Helikites since its deployment can take up to 45 minutes. In this case the DBSs will heal the users until the Helikite is deployed and then the DBSs will return back to its initial location.

\subsection{Numerical Results}

Numerical results are provided to investigate the benefits of using different types of UAVs to mitigate GBSs failure using D$^3$S framework. 

The optimization problem presented in \cite{8648054} is solved using General Algebraic Modeling System (GAMS) (https://www.gams.com/). The simulation area is 400x400 m$^2$ and the UEs are distributed randomly. The parameters used in the simulation can be found in \cite{8648054}.

\begin{figure}
    \centering
    \includegraphics[width=90mm]{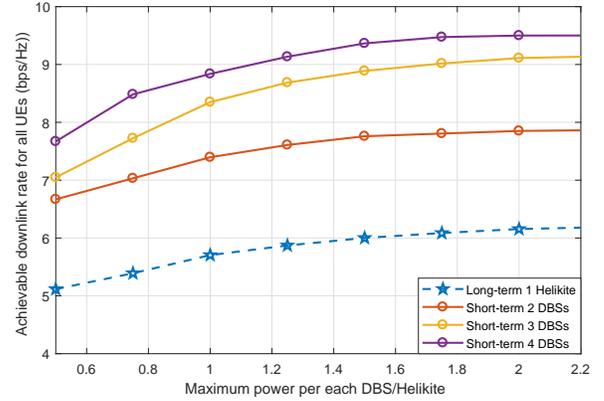}
    \caption{Achievable downlink rate for all UEs.}
    \label{fig:rates}
\end{figure}

Fig.~\ref{fig:rates} represents failure mitigation performance for short-term and long-term failures in terms of the achievable downlink rate. By increasing the number of used DBSs, the consumed power increases. As the maximum power increases, the rate increases but levels off when the power reaches 1W. Owing to the fact that the objective function of the optimization problem is maximizing the minimum achievable rate and at the same time minimizing the downlink power.

The long-term scenario which uses one Helikite results in the lowest achievable rate. The reason for that is the altitude of the Helikite is higher than that for the drones.

Table ~\ref{tabelx} shows the UAV-UE association and UEs power for short-term and long-term scenarios. For long-term failure, the maximum power assigned to the Helikite is 2.25 W. Since in this scenario only one Helikite is used, a variety in power levels among different UEs is observed. For example, UE3 has the least power consumption and this means it is close to the Helikite given that the free space loss model is used for the path-loss. Furthermore, UE2 and UE8 use around 40\% of the Helikite maximum power since the Helikite is covering the whole area of the failed GBS, hence, satisfying the minimum rate of the far located UEs by increasing their transmission power.

For short-term failure, there are 4 DBSs available/standby, however, only three DBSs are used as shown in Table ~\ref{tabelx}. It worth noting that DBS1 and DBS2 utilize less than 50\% of their maximum power since in this scenario not all UEs are associated with one DBS. On the contrary, DBS4 utilized around 95\% of its maximum power. This is because more than three UEs are connected to DBS4.

\small
\begin{table}
  \centering
  \vspace*{0.15in}
  \caption{\label{tab3} Association and power for 10 UEs}
  \label{tabelx}
  \renewcommand{\arraystretch}{1.2}
  \begin{tabular}{|p{1cm}|c|c|c|c|}
    \hline
    %\multirow{\textbf{UEs}} & \multicolumn{2}{c|}{Short-term} & \multicolumn{2}{c|}{Long-term}\\
    % \hline
    \cline{2-5}
%    & \textbf{Association} & {$p_{u,d}$ (W)} & \textbf{Association} & {$p_{u,d}$ (W)} \\
    %\hhline{~--}
    & \textbf{Association} & Power & \textbf{Association} & Power \\
    %\hhline{~--}
    \hline
    UE1 & GBS4 & 0.156 & Helikite  &0.176\\ \hline
    UE2 & GBS1 & 0.147 & Helikite  &0.397 \\ \hline
    UE3 & DBS4 & 0.105 & Helikite  &0.108 \\ \hline
    UE4 & GBS2 & 0.197 & Helikite  &0.203\\ \hline
    UE5 & DBS4 & 0.130 & Helikite  &0.239\\ \hline
    UE6 & GBS1 & 0.132 & Helikite  &0.115 \\ \hline
    UE7 & GBS4 & 0.171 & Helikite  &0.279 \\ \hline
    UE8 & GBS2 & 0.121 & Helikite  &0.451\\ \hline
    UE9 & DBS4 & 0.164 & Helikite  &0.153 \\ \hline
    UE10& GBS1 & 0.139 & Helikite  &0.129\\ \hline
  \end{tabular}
\vspace{-0.6cm}
\end{table}
\normalsize 
%\subsection{Case Study II: UaaS for Disaster Recovery}

\section{Conclusion}
In this article, we introduce a novel framework of UAVs as a Service (UaaS) and showcase its usage in the context of wireless  connectivity service. 
Based on four phases (\textit{Demand}, \textit{Decision}, \textit{Deployment} and \textit{Service}), the main objectives of this framework is to develop efficient and realistic solutions to implement these four phases.
To evaluate the performance of this framework, we illustrate its application in a case study that addresses a failure of one or more Ground BSs (GBSs) of a wireless cellular network and show how we can mitigate the effect of this failure to keep the wireless connectivity service operational.

\bibliographystyle{IEEEtran}  
\bibliography{./bibliographies/main-d3s-uav-commag.bib}

\end{document}